\documentclass[a4paper,11pt]{article}
\usepackage[latin1]{inputenc}
\usepackage{amssymb}

\usepackage{times}

\title{{\bf Some Problems  in Automata Theory  \\ Which Depend on  the  Models of  Set Theory }} 
\author{Olivier Finkel \\
{\it Equipe de Logique Math\'ematique}\\ Institut de Math\'ematiques de Jussieu
 \\  CNRS et Universit\'e Paris 7, France. \\ 
finkel@logique.jussieu.fr }

\date{}
\begin{document}

\newtheorem{theorem}{Theorem}[section]
\newtheorem{Rem}[theorem]{Remark}
\newtheorem{Exa}[theorem]{Example}

\newtheorem{Pro}[theorem]{Proposition}
\newtheorem{lem}[theorem]{Lemma}
\newtheorem{Cor}[theorem]{Corollary}
\newtheorem{defi}[theorem]{Definition}
\newtheorem{notation}[theorem]{Notation}

\def\ufootnote#1{\let\savedthfn\thefootnote\let\thefootnote\relax
\footnote{#1}\let\thefootnote\savedthfn\addtocounter{footnote}{-1}}

\newcommand{\bormxi}{{\bf\Pi}^{0}_{\xi}}
\newcommand{\bormlxi}{{\bf\Pi}^{0}_{<\xi}}
\newcommand{\bormz}{{\bf\Pi}^{0}_{0}}
\newcommand{\bormone}{{\bf\Pi}^{0}_{1}}
\newcommand{\ca}{{\bf\Pi}^{1}_{1}}
\newcommand{\bormtwo}{{\bf\Pi}^{0}_{2}}
\newcommand{\bormthree}{{\bf\Pi}^{0}_{3}}
\newcommand{\bormom}{{\bf\Pi}^{0}_{\omega}}
\newcommand{\borom}{{\bf\Delta}^{0}_{\omega}}
\newcommand{\borml}{{\bf\Pi}^{0}_{\lambda}}
\newcommand{\bormlpn}{{\bf\Pi}^{0}_{\lambda +n}}
\newcommand{\bormpm}{{\bf\Pi}^{0}_{1+m}}
\newcommand{\borapm}{{\bf\Sigma}^{0}_{1+m}}
\newcommand{\bormep}{{\bf\Pi}^{0}_{\eta +1}}
\newcommand{\borapxi}{{\bf\Sigma}^{0}_{\xi}}
\newcommand{\borai}{{\bf\Sigma}^{0}_{ 2.\xi +1 }}
\newcommand{\bormpxi}{{\bf\Pi}^{0}_{\xi}}
\newcommand{\bormpeta}{{\bf\Pi}^{0}_{1+\eta}}
\newcommand{\borapxipo}{{\bf\Sigma}^{0}_{\xi +1}}
\newcommand{\bormpxipo}{{\bf\Pi}^{0}_{\xi +1}}
\newcommand{\borpxi}{{\bf\Delta}^{0}_{\xi}}
\newcommand{\borel}{{\bf\Delta}^{1}_{1}}
\newcommand{\Borel}{{\it\Delta}^{1}_{1}}
\newcommand{\borone}{{\bf\Delta}^{0}_{1}}
\newcommand{\bortwo}{{\bf\Delta}^{0}_{2}}
\newcommand{\borthree}{{\bf\Delta}^{0}_{3}}
\newcommand{\boraone}{{\bf\Sigma}^{0}_{1}}
\newcommand{\boratwo}{{\bf\Sigma}^{0}_{2}}
\newcommand{\borathree}{{\bf\Sigma}^{0}_{3}}
\newcommand{\boraom}{{\bf\Sigma}^{0}_{\omega}}
\newcommand{\boraxi}{{\bf\Sigma}^{0}_{\xi}}
\newcommand{\ana}{{\bf\Sigma}^{1}_{1}}
\newcommand{\pca}{{\bf\Sigma}^{1}_{2}}
\newcommand{\Ana}{{\it\Sigma}^{1}_{1}}
\newcommand{\Boraone}{{\it\Sigma}^{0}_{1}}
\newcommand{\Borone}{{\it\Delta}^{0}_{1}}
\newcommand{\Bormone}{{\it\Pi}^{0}_{1}}
\newcommand{\Bormtwo}{{\it\Pi}^{0}_{2}}
\newcommand{\Ca}{{\it\Pi}^{1}_{1}}
\newcommand{\bormn}{{\bf\Pi}^{0}_{n}}
\newcommand{\bormm}{{\bf\Pi}^{0}_{m}}
\newcommand{\boralp}{{\bf\Sigma}^{0}_{\lambda +1}}
\newcommand{\borat}{{\bf\Sigma}^{0}_{|\theta |}}
\newcommand{\bormat}{{\bf\Pi}^{0}_{|\theta |}}
\newcommand{\Borapxi}{{\it\Sigma}^{0}_{\xi}}
\newcommand{\Bormpxipo}{{\it\Pi}^{0}_{1+\xi +1}}
\newcommand{\Borapn}{{\it\Sigma}^{0}_{1+n}}
\newcommand{\borapn}{{\bf\Sigma}^{0}_{1+n}}
\newcommand{\boraxipm}{{\bf\Sigma}^{0}_{\xi^\pm}}
\newcommand{\Boratwo}{{\it\Sigma}^{0}_{2}}
\newcommand{\Borathree}{{\it\Sigma}^{0}_{3}}
\newcommand{\Borapnpo}{{\it\Sigma}^{0}_{1+n+1}}
\newcommand{\Bormpxi}{{\it\Pi}^{0}_{\xi}}
\newcommand{\Borpxi}{{\it\Delta}^{0}_{\xi}}
\newcommand{\boratpxi}{{\bf\Sigma}^{0}_{2+\xi}}
\newcommand{\Boratpxi}{{\it\Sigma}^{0}_{2+\xi}}
\newcommand{\bormltpxi}{{\bf\Pi}^{0}_{<2+\xi}}
\newcommand{\Bormltpxi}{{\it\Pi}^{0}_{<2+\xi}}
\newcommand{\borapeap}{{\bf\Sigma}^{0}_{1+\eta_{\alpha ,p}}}
\newcommand{\borapeapn}{{\bf\Sigma}^{0}_{1+\eta_{\alpha ,p,n}}}
\newcommand{\Borapeap}{{\it\Sigma}^{0}_{1+\eta_{\alpha ,p}}}
\newcommand{\Bormpn}{{\it\Pi}^{0}_{1+n}}
\newcommand{\Borpn}{{\it\Delta}^{0}_{1+n}}
\newcommand{\borapximo}{{\bf\Sigma}^{0}_{1+(\xi -1)}}
\newcommand{\borpeta}{{\bf\Delta}^{0}_{1+\eta}}

\newcommand{\hs}{\hspace{12mm}

}
\newcommand{\noi}{\noindent}

\newcommand{\om}{\omega}
\newcommand{\Si}{\Sigma}
\newcommand{\Sis}{\Sigma^\star}
\newcommand{\Sio}{\Sigma^\omega}
\newcommand{\nl}{\newline}
\newcommand{\lra}{\leftrightarrow}
\newcommand{\fa}{\forall}
\newcommand{\ra}{\rightarrow}
\newcommand{\orl}{ $\omega$-regular language}

\newcommand{\Ga}{\Gamma}
\newcommand{\Gas}{\Gamma^\star}
\newcommand{\Gao}{\Gamma^\omega}
\newcommand{\ite}{\item}
\newcommand{\la}{language}
\newcommand{\Lp}{L(\varphi)}
\newcommand{\abs}{\{a, b\}^\star}
\newcommand{\abcs}{\{a, b, c \}^\star}
\newcommand{\ol}{$\omega$-language}

\newcommand{\tla}{\twoheadleftarrow}
\newcommand{\de}{deterministic }
\newcommand{\proo}{\noi {\bf Proof.} }
\newcommand {\ep}{\hfill $\square$}

\maketitle

\begin{abstract}
\noi We prove that  some fairly basic questions on automata reading infinite words depend on the models of the axiomatic system 
{\bf ZFC}.  It is known that there are only  three possibilities for the cardinality of the
 complement  of an $\om$-language $L(\mathcal{A})$ accepted by a B\"uchi $1$-counter automaton $\mathcal{A}$. 
We prove the following  surprising result: there  exists a $1$-counter 
B\"uchi automaton $\mathcal{A}$ 
 such that the  cardinality of the complement $L(\mathcal{A})^-$ of the 
$\om$-language $L(\mathcal{A})$ is not determined by {\bf ZFC}: 

 (1). 
There is a model $V_1$ of  {\bf ZFC} in which      $L(\mathcal{A})^-$ is countable. 

(2).  There is a model $V_2$ of  {\bf ZFC} in which     $L(\mathcal{A})^-$ has cardinal $2^{\aleph_0}$.

(3). There is a model $V_3$ of  {\bf ZFC} in which      $L(\mathcal{A})^-$ has cardinal $\aleph_1$ with 
 $\aleph_0<\aleph_1<2^{\aleph_0}$.

\noi 
We prove a very similar result for the  complement of  an   infinitary rational relation accepted by a 
$2$-tape B\"uchi automaton  $\mathcal{B}$. 
 As a corollary, this proves that the Continuum Hypothesis may be not satisfied for complements of $1$-counter 
 $\om$-languages and  for complements of  infinitary rational relations accepted by $2$-tape B\"uchi automata. 
\nl We infer from the proof of the above results that  basic decision problems about  $1$-counter 
 $\om$-languages or  infinitary rational relations are actually located at the {\bf third level} of the analytical hierarchy. In particular,  the problem 
to determine whether the complement of a $1$-counter $\om$-language (respectively, infinitary rational relation) is countable is 
in  $\Si_3^1 \setminus (\Pi_2^1 \cup \Si_2^1)$. 
This is rather surprising if compared to the fact   that it is  {\bf decidable}  whether an infinitary rational relation is countable  (respectively, uncountable). 

\end{abstract}


\section{Introduction}

~~~~~ In Computer Science one usually considers   either finite computations or infinite ones. 
The infinite computations have length $\om$, which is the first 
infinite ordinal. The theory of automata reading infinite words, which is closely related to infinite games,  is now a rich theory which is used for the 
specification and verification of non-terminating systems, see \cite{2001automata,PerrinPin}. 

Connections between Automata Theory and Set Theory have arosen in the study of monadic theories of well orders. 
For example, Gurevich, Magidor and Shelah proved in \cite{GMS83} that the monadic theory of $\om_2$, where $\om_2$ is the second uncountable cardinal, 
may have different complexities depending on the actual model of  {\bf ZFC} (the commonly accepted axiomatic 
framework for Set Theory in which all usual 
mathematics can be developped), and the monadic theory of $\om_2$ is in turn 
closely related to the emptiness problem for automata reading transfinite words 
of length  $\om_2$. 
Another example is given by   \cite{Neeman08}, in which  Neeman considered 
 some automata reading much longer transfinite words to study the monadic theory of some larger
 uncountable cardinal. 

 However, the cardinal $\om_2$ is  very large with respect  to $\om$, and therefore 
 the  connections between Automata Theory and Set Theory seemed very far from 
the practical aspects of Computer Science. Indeed one usually thinks   that the finite or infinite computations appearing in Computer Science 
are ``well defined" in the axiomatic 
framework of  mathematics, and thus one could be tempted to consider  that a property on automata is either true or false  
and that one has not to take care of  the different models of Set Theory 
(except perhaps for the Continuum Hypothesis {\bf CH} which is known to be independent from {\bf ZFC}). 

 In \cite{Fin-ICST} we have recently proved   a surprising result:  
 the topological complexity of an $\om$-language accepted by a $1$-counter B\"uchi automaton, or of an infinitary rational relation 
accepted by a $2$-tape B\"uchi automaton, 
is not determined by the  axiomatic system {\bf ZFC}. In particular, there is a $1$-counter B\"uchi   automaton $\mathcal{A}$ 
(respectively, a $2$-tape B\"uchi   automaton   $\mathcal{B}$)   and two models ${\bf V}_1$ and ${\bf V}_2$
of {\bf ZFC} such that  the $\om$-language $L(\mathcal{A})$ (respectively, the infinitary rational relation     $L(\mathcal{B})$)    
 is Borel in ${\bf V}_1$ but not  in ${\bf V}_2$. 

We prove in this paper other surprising results, showing  that some 
basic questions on automata reading infinite words actually depend on the models of  
{\bf ZFC}. 
In particular, we prove the following   result: there  exists a $1$-counter 
B\"uchi automaton $\mathcal{A}$ 
 such that the  cardinality of the complement $L(\mathcal{A})^-$ of the 
$\om$-language $L(\mathcal{A})$ is not determined by  {\bf ZFC}. Indeed it holds that: 

(1).  There is a model $V_1$ of  {\bf ZFC} in which      $L(\mathcal{A})^-$ is countable. 

(2).  There is a model $V_2$ of  {\bf ZFC} in which     $L(\mathcal{A})^-$ has cardinal $2^{\aleph_0}$. 

(3).  There is a model $V_3$ of  {\bf ZFC} in which      $L(\mathcal{A})^-$ has cardinal $\aleph_1$ with 
 $\aleph_0<\aleph_1<2^{\aleph_0}$.
 
Notice that  there are only these  three possibilities for the cardinality of the
 complement  of an $\om$-language  accepted by a B\"uchi $1$-counter automaton $\mathcal{A}$ because the $\om$-language $L(\mathcal{A})$ is an analytic set 
and thus  $L(\mathcal{A})^-$ is a coanalytic set, see  
\cite[page 488]{Jech}. 

 We prove a very similar result for the  complement of  an   infinitary rational relation accepted by a 
$2$-tape B\"uchi automaton  $\mathcal{B}$. 
 As a corollary, this proves that the Continuum Hypothesis may be not satisfied for complements of $1$-counter 
 $\om$-languages and  for complements of  infinitary rational relations accepted by $2$-tape B\"uchi automata. 

In the proof of  these results, we consider the largest thin (i.e., without perfect subset) effective coanalytic subset of the Cantor space $2^\om$, whose  
 existence was proven by Kechris in \cite{Kechris75} and independently 
by Guaspari  and  Sacks. An important property of  $\mathcal{C}_1$   is that its cardinal depends on the  models of set theory. We use this fact  
and  some constructions from recent papers 
\cite{Fin-mscs06,Fin06b} to infer our new results about $1$-counter or 
$2$-tape  B\"uchi   automata.

Combining  the proof of the above results with  Shoenfield's Absoluteness Theorem we get that  basic decision problems about  $1$-counter 
 $\om$-languages or  infinitary rational relations are actually located at the {\bf third level} of the analytical hierarchy. 
In particular,  the problem 
to determine whether the complement of a $1$-counter $\om$-language (respectively, infinitary rational relation) is countable is 
in  $\Si_3^1 \setminus (\Pi_2^1 \cup \Si_2^1)$. 
This is rather surprising if compared to the fact   that it is  {\bf decidable}  whether an infinitary rational relation is countable  (respectively, uncountable). 
As a by-product of these results we get a (partial) answer to a question of Castro and Cucker about $\om$-languages of Turing machines. 

The paper is organized as follows. We recall the notion of counter automata in Section 2. We expose some results of Set Theory in Section 3, 
and we prove our main results in Section 4. Concluding remarks are given in Section 5. 

\hs Notice that the reader who is not familiar  with the notion of ordinal  in set theory may  skip part of Section 3 and just  read  
Theorems \ref{thin} and \ref{cor1} in this section. The rest of the paper relies mainly on the set-theoretical results  stated in Theorem  \ref{cor1}. 

\section{Counter Automata}
 
~~~~~  We assume   the reader to be familiar with the theory of formal ($\om$-)languages  
\cite{Thomas90,Staiger97}.
We recall the  usual notations of formal language theory. 

If  $\Si$ is a finite alphabet, a {\it non-empty finite word} over $\Si$ is any 
sequence $x=a_1\ldots a_k$, where $a_i\in\Sigma$ 
for $i=1,\ldots ,k$ , and  $k$ is an integer $\geq 1$. The {\it length}
 of $x$ is $k$, denoted by $|x|$.
 The {\it empty word} has no letter and is denoted by $\lambda$; its length is $0$. 
 $\Sis$  is the {\it set of finite words} (including the empty word) over $\Sigma$.
 
 The {\it first infinite ordinal} is $\om$.
 An $\om$-{\it word} over $\Si$ is an $\om$ -sequence $a_1 \ldots a_n \ldots$, where for all 
integers $ i\geq 1$, ~
$a_i \in\Sigma$.  When $\sigma=a_1 \ldots a_n \ldots$ is an $\om$-word over $\Si$, we write
 $\sigma(n)=a_n$,   $\sigma[n]=\sigma(1)\sigma(2)\ldots \sigma(n)$  for all $n\geq 1$ and $\sigma[0]=\lambda$.

 The usual concatenation product of two finite words $u$ and $v$ is 
denoted $u.v$ (and sometimes just $uv$). This product is extended to the product of a 
finite word $u$ and an $\om$-word $v$: the infinite word $u.v$ is then the $\om$-word such that:

 $(u.v)(k)=u(k)$  if $k\leq |u|$ , and 
 $(u.v)(k)=v(k-|u|)$  if $k>|u|$.
  
 The {\it set of } $\om$-{\it words} over  the alphabet $\Si$ is denoted by $\Si^\om$.
An  $\om$-{\it language} $V$ over an alphabet $\Sigma$ is a subset of  $\Si^\om$, and its  complement (in $\Sio$) 
 is $\Sio - V$, denoted $V^-$.

 \hs We now recall the definition of $k$-counter B\"uchi automata which will be useful in the sequel. 

 Let $k$ be an integer $\geq 1$. 
A  $k$-counter machine has $k$ {\it counters}, each of which containing a  non-negative integer. 
The machine can test whether the content of a given counter is zero or not. 
And transitions depend on the letter read by the machine, the current state of the finite control, and the tests about the values of the counters. 
Notice that in this model some  $\lambda$-transitions are allowed. During these transitions the reading head of the machine does not move to the right, i.e. 
 the machine does not  read any more letter. 

Formally a  $k$-counter machine is a 4-tuple 
$\mathcal{M}$=$(K,\Si,$ $ \Delta, q_0)$,  where $K$ 
is a finite set of states, $\Sigma$ is a finite input alphabet, 
 $q_0\in K$ is the initial state, 
and  $\Delta \subseteq K \times ( \Si \cup \{\lambda\} ) \times \{0, 1\}^k \times K \times \{0, 1, -1\}^k$ is the transition relation. 
The $k$-counter machine $\mathcal{M}$ is said to be {\it real time} iff: 
$\Delta \subseteq K \times
  \Si \times \{0, 1\}^k \times K \times \{0, 1, -1\}^k$, 
 i.e. iff there are no  $\lambda$-transitions. 

If  the machine $\mathcal{M}$ is in state $q$ and 
$c_i \in \mathbf{N}$ is the content of the $i^{th}$ counter 
 $\mathcal{C}$$_i$ then 
the  configuration (or global state)
 of $\mathcal{M}$ is the  $(k+1)$-tuple $(q, c_1, \ldots , c_k)$.

 For $a\in \Si \cup \{\lambda\}$, 
$q, q' \in K$ and $(c_1, \ldots , c_k) \in \mathbf{N}^k$ such 
that $c_j=0$ for $j\in E \subseteq  \{1, \ldots , k\}$ and $c_j >0$ for 
$j\notin E$, if 
$(q, a, i_1, \ldots , i_k, q', j_1, \ldots , j_k) \in \Delta$ where $i_j=0$ for $j\in E$ 
and $i_j=1$ for $j\notin E$, then we write:

~~~~~~~~$a: (q, c_1, \ldots , c_k)\mapsto_{\mathcal{M}} (q', c_1+j_1, \ldots , c_k+j_k)$.

 Thus  the transition relation must obviously satisfy:
 \nl if $(q, a, i_1, \ldots , i_k, q', j_1, \ldots , j_k)  \in    \Delta$ and  $i_m=0$ for 
 some $m\in \{1, \ldots , k\}$  then $j_m=0$ or $j_m=1$ (but $j_m$ may not be equal to $-1$). 
  
Let $\sigma =a_1a_2 \ldots a_n \ldots $ be an $\om$-word over $\Si$. 
An $\om$-sequence of configurations $r=(q_i, c_1^{i}, \ldots c_k^{i})_{i \geq 1}$ is called 
a run of $\mathcal{M}$ on $\sigma$, starting in configuration 
$(p, c_1, \ldots, c_k)$, iff:

(1)  $(q_1, c_1^{1}, \ldots c_k^{1})=(p, c_1, \ldots, c_k)$

(2)   for each $i\geq 1$, there  exists $b_i \in \Si \cup \{\lambda\}$ such that
 $b_i: (q_i, c_1^{i}, \ldots c_k^{i})\mapsto_{\mathcal{M}}  
(q_{i+1},  c_1^{i+1}, \ldots c_k^{i+1})$  
and such that either ~  $a_1a_2\ldots a_n\ldots =b_1b_2\ldots b_n\ldots$ 
\nl or ~  $b_1b_2\ldots b_n\ldots$ is a finite word,  prefix (i.e. initial segment) of ~ $a_1a_2\ldots a_n\ldots$

The run $r$ is said to be complete when $a_1a_2\ldots a_n\ldots =b_1b_2\ldots b_n\ldots$ 

For every such run $r$, $\mathrm{In}(r)$ is the set of all states entered infinitely
 often during $r$.

A complete run $r$ of $M$ on $\sigma$, starting in configuration $(q_0, 0, \ldots, 0)$,
 will be simply called ``a run of $M$ on $\sigma$".

\begin{defi} A B\"uchi $k$-counter automaton  is a 5-tuple 
$\mathcal{M}$=$(K,\Si, \Delta, q_0, F)$, 
where $ \mathcal{M}'$=$(K,\Si, \Delta, q_0)$
is a $k$-counter machine and $F \subseteq K$ 
is the set of accepting  states.
The \ol~ accepted by $\mathcal{M}$ is:~~ $L(\mathcal{M})$= $\{  \sigma\in\Si^\om \mid \mbox{  there exists a  run r
 of } \mathcal{M} \mbox{ on } \sigma \mbox{  such that } \mathrm{In}(r)
 \cap F \neq \emptyset \}$

\end{defi}

  The class of \ol s accepted by  B\"uchi $k$-counter automata  is  
denoted ${\bf BCL}(k)_\om$.
 The class of \ol s accepted by {\it  real time} B\"uchi $k$-counter automata  will be 
denoted {\bf r}-${\bf BCL}(k)_\om$.
  The class ${\bf BCL}(1)_\om$ is  a strict subclass of the class ${\bf CFL}_\om$ of context free \ol s
accepted by B\"uchi pushdown automata.

\hs We recall now   the definition of classes of the arithmetical hierarchy of $\om$-languages, see \cite{Staiger97}. 
Let $X$ be a finite alphabet. An \ol~ $L\subseteq X^\om$  belongs to the class 
$\Si_n$ if and only if there exists a recursive relation 
$R_L\subseteq (\mathbb{N})^{n-1}\times X^\star$  such that:
$$L = \{\sigma \in X^\om \mid \exists a_1\ldots Q_na_n  \quad (a_1,\ldots , a_{n-1}, 
\sigma[a_n+1])\in R_L \},$$
\noi where $Q_i$ is one of the quantifiers $\fa$ or $\exists$ 
(not necessarily in an alternating order). An \ol~ $L\subseteq X^\om$  belongs to the class 
$\Pi_n$ if and only if its complement $X^\om - L$  belongs to the class 
$\Si_n$.  
The  class  $\Si^1_1$ is the class of {\it effective analytic sets} which are 
 obtained by projection of arithmetical sets.
An \ol~ $L\subseteq X^\om$  belongs to the class 
$\Si_1^1$ if and only if there exists a recursive relation 
$R_L\subseteq \mathbb{N}\times \{0, 1\}^\star \times X^\star$  such that:
$$L = \{\sigma \in X^\om  \mid \exists \tau (\tau\in \{0, 1\}^\om \wedge \fa n \exists m 
 ( (n, \tau[m], \sigma[m]) \in R_L )) \}.$$
\noi 
 Then an \ol~ $L\subseteq X^\om$  is in the class $\Si_1^1$ iff it is the projection 
of an \ol~ over the alphabet $X\times \{0, 1\}$ which is in the class $\Pi_2$.  The   class $\Pi_1^1$ of  {\it effective co-analytic sets} 
 is simply the class of complements of effective analytic sets. 

\hs  Recall that a B\"uchi Turing machine is just a Turing machine working on infinite inputs with a B\"uchi-like acceptance condition, and 
that the class of  $\om$-languages accepted by  B\"uchi Turing machines is the class $ \Si^1_1$ of effective analytic sets  \cite{CG78b,Staiger97}.
On the oher hand, one can  construct, using a classical  construction (see for instance  \cite{HopcroftMotwaniUllman2001}),  from a  B\"uchi 
Turing machine $\mathcal{T}$,  a $2$-counter  B\"uchi  automaton $\mathcal{A}$ accepting the same $\om$-language. 
Thus one can state the following proposition. 

\begin{Pro}\label{tm}
An \ol~ $L\subseteq X^\om$ is in the class $\Si_1^1$
iff it is accepted by a non deterministic  B\"uchi  Turing machine,  hence  iff it is in the class ${\bf BCL}(2)_\om$. 
\end{Pro}

 \section{Some Results  of Set Theory}

~~~~~~ 
We recall  that the reader who is not familiar  with the notion of ordinal  in set theory may  skip part of this section: 
 the main results in this section, which will be used later in this paper, are  stated in Theorems \ref{thin} and   \ref{cor1}. 

 We now recall some basic notions of set theory 
which will be useful in the sequel, and which are exposed in any  textbook on set theory, like \cite{Jech}. 

 The usual axiomatic system {\bf ZFC} is 
Zermelo-Fraenkel system {\bf ZF}   plus the axiom of choice {\bf AC}. 
 The axioms of {\bf ZFC} express some  natural facts that we consider to hold in the universe of sets. For instance a natural fact is that 
two sets $x$ and $y$ are equal iff they have the same elements. 
This is expressed by the {\it Axiom of Extensionality}: 
$$\fa x \fa y ~ [ ~ x=y \leftrightarrow \fa z ( z\in x \leftrightarrow z\in y ) ~].$$
\noi   Another natural axiom is the {\it Pairing Axiom}   which states that for all sets $x$ and $y$ there exists a  set $z=\{x, y\}$ 
whose elements are $x$ and $y$: 
 $$\fa x \fa y ~ [ ~\exists z ( \fa w  ( w\in z \leftrightarrow (w=x \vee w=y) ) ) ]$$
\noi Similarly the {\it Powerset Axiom} states the existence of the set of subsets of a set $x$. 
Notice that these axioms are first-order sentences in the usual logical language of set theory whose only non logical  
symbol is the membership binary relation symbol $\in$. 
We refer the reader to any textbook on set theory  for an exposition of the other axioms of {\bf ZFC}.  

A model ({\bf V}, $\in)$ of  an arbitrary set of axioms $\mathbb{A}$  is a collection  {\bf V} of sets,  equipped with 
the membership relation $\in$, where ``$x \in y$" means that the set $x$ is an element of the set $y$, which satisfies the axioms of   $\mathbb{A}$. 
We  often say `` the model {\bf V}" instead of  `` the model ({\bf V}, $\in)$".

We say that two sets $A$ and $B$ have same cardinality iff there is a bijection from $A$ onto $B$ and we denote this  by $A \approx B$. 
The relation $\approx$ is an equivalence relation. 
Using the axiom of choice {\bf AC}, one can prove that any set $A$ can be well-ordered so  there is an ordinal $\gamma$ such that $A \approx \gamma$. 
In set theory the cardinal of the set $A$ is then formally defined as the smallest such ordinal $\gamma$.

 The infinite cardinals are usually denoted by
$\aleph_0, \aleph_1, \aleph_2, \ldots , \aleph_\alpha, \ldots$
The cardinal $\aleph_\alpha$ is also denoted by $\om_\alpha$,
 when it is considered as an ordinal.
The first infinite ordinal is $\om$ and it is the smallest ordinal which is countably infinite so $\aleph_0=\om$ (which could  be written $\om_0$). 
There are many larger countable ordinals, such as $\om^2, \om^3, \ldots , \om^\om, \ldots \om^{\om^\om}, \ldots$ The first uncountable ordinal is $\om_1$, 
 and formally  $\aleph_1=\om_1$. In the same way $\om_2$ is the first ordinal of cardinality 
greater than $\aleph_1$, and so on. 

 The continuum hypothesis {\bf CH}  says that the first uncountable cardinal $\aleph_1$ is equal to $2^{\aleph_0}$ which is the cardinal of the 
continuum. 
G\"odel and Cohen have proved that the continuum hypothesis {\bf CH} is independent from the axiomatic system {\bf ZFC}, i.e.,  
 that there are models of {\bf ZFC + CH} and also  models of {\bf ZFC + $\neg$ CH}, where {\bf $\neg$ CH} denotes the negation of the 
continuum hypothesis, \cite{Jech}.

Let ${\bf ON}$ be the class of all ordinals. Recall that an ordinal $\alpha$ is said to be a successor ordinal iff there exists an ordinal $\beta$ such that 
$\alpha=\beta + 1$; otherwise the ordinal $\alpha$ is said to be a limit ordinal and in this case 
$\alpha ={\rm sup} \{ \beta \in {\bf ON}\mid \beta < \alpha \}$.

\hs  The  class ${\bf L}$ of  {\it constructible sets} in a model {\bf V} of {\bf ZF} is defined by ~~~~
${\bf L} = \bigcup_{\alpha \in {\bf ON}} {\bf L}(\alpha) $, 
\noi where the sets ${\bf L}(\alpha) $ are constructed  by induction as follows: 

(1). ${\bf L}(0) =\emptyset$

(2). ${\bf L}(\alpha) = \bigcup_{\beta <  \alpha} {\bf L}(\beta) $, for $\alpha$ a limit ordinal, and 

(3).  ${\bf L}(\alpha + 1) $ is the set of subsets of ${\bf L}(\alpha) $ which are definable from a finite number of elements of ${\bf L}(\alpha) $
by a first-order formula relativized to ${\bf L}(\alpha) $.

  If  {\bf V} is  a model of {\bf ZF} and ${\bf L}$ is  the class of  {\it constructible sets} of   {\bf V}, then the class  ${\bf L}$    is a model of  
{\bf ZFC + CH}.
Notice that the axiom ({\bf V=L}), which  means ``every set is constructible",   is consistent with {\bf ZFC}  because   ${\bf L}$ is a model of 
{\bf ZFC + V=L}. 

 Consider now a model {\bf V} of  {\bf ZFC} and the class of its constructible sets ${\bf L} \subseteq {\bf V}$ which is another 
model of  {\bf ZFC}.  It is known that 
the ordinals of {\bf L} are also the ordinals of  {\bf V}, but the cardinals  in  {\bf V}  may be different from the cardinals in {\bf L}. 

  In particular,  the first uncountable cardinal in {\bf L}  is denoted 
 $\aleph_1^{\bf L}$, and it is in fact an ordinal of {\bf V} which is denoted $\om_1^{\bf L}$. 
  It is well-known that in general this ordinal satisfies the inequality 
$\om_1^{\bf L} \leq \om_1$.  In a model {\bf V} of  the axiomatic system {\bf ZFC + V=L} the equality $\om_1^{\bf L} = \om_1$ holds, but in 
some other models of {\bf ZFC} the inequality may be strict and then $\om_1^{\bf L} < \om_1$: 
notice that in this case $\om_1^{\bf L} < \om_1$ holds because there is actually a bijection from $\om$ onto $\om_1^{\bf L}$ in {\bf V} 
(so  $\om_1^{\bf L}$ is countable in {\bf V})  but no such bijection exists in the inner model {\bf L} (so $\om_1^{\bf L}$ is uncountable in {\bf L}). 
The construction of such a model is presented in \cite[page 202]{Jech}: one can start 
 from a model 
{\bf V} of {\bf ZFC + V=L} and construct by  forcing  a generic extension {\bf V[G]} in which 
$\om_1^{{\bf V}}$ is collapsed to 
 $\om$; in this extension the inequality $\om_1^{\bf L} < \om_1$ holds. 

\hs We assume the reader to be familiar with basic notions of topology which
may be found in \cite{Moschovakis80,LescowThomas,Staiger97,PerrinPin}.
There is a natural metric on the set $\Sio$ of  infinite words 
over a finite alphabet 
$\Si$ containing at least two letters which is called the {\it prefix metric} and is defined as follows. For $u, v \in \Sio$ and 
$u\neq v$ let $\delta(u, v)=2^{-l_{\mathrm{pref}(u,v)}}$ where $l_{\mathrm{pref}(u,v)}$ 
 is the first integer $n$
such that the $(n+1)^{st}$ letter of $u$ is different from the $(n+1)^{st}$ letter of $v$. 
This metric induces on $\Sio$ the usual  Cantor topology in which the {\it open subsets} of 
$\Sio$ are of the form $W.\Si^\om$, for $W\subseteq \Sis$.
A set $L\subseteq \Si^\om$ is a {\it closed set} iff its complement $\Si^\om - L$ 
is an open set.

\begin{defi} 
Let $P \subseteq \Sio$, where $\Si$ is a finite alphabet having at least two letters. The set 
 $P$ is said to be a perfect subset of $\Sio$ if and only if :  
\nl  (1) $P$ is a non-empty closed set,  and 
\nl (2) for every $x\in P$ and every open set $U$ containing $x$ there is an element 
$y \in P\cap U$ such that $x\neq y$. 
\end{defi}

 So a perfect subset of $\Sio$ is a non-empty closed set which has no isolated points. It is well known that a  perfect subset of 
$\Sio$   has cardinality $2^{\aleph_0}$, i.e. the cardinality of the continuum, see \cite[page 66]{Moschovakis80}.

 \begin{defi}
A set $X \subseteq \Sio$ is said to be thin iff it  contains no perfect subset. 
\end{defi}

 The  following result was proved by Kechris \cite{Kechris75} and independently by Guaspari  and  Sacks. 

\begin{theorem}[see \cite{Moschovakis80} page 247]\label{thin}  ({\bf ZFC})  Let $\Si$ be a finite alphabet having at least two letters. 
There exists a thin $\Pi_1^1$-set $\mathcal{C}_1( \Sio) \subseteq  \Sio$ which contains every thin,  $\Pi_1^1$-subset of $\Sio$. 
It is called the  largest thin $\Pi_1^1$-set  in $\Sio$.    
\end{theorem}

 An important fact is that the cardinality of the largest thin $\Pi_1^1$-set in $\Sio$  depends on the model of {\bf ZFC}. 
The following result was proved by Kechris, and  independently by Guaspari and Sacks, see  
\cite[page 171]{Kanamori}. 

\begin{theorem}\label{card}
({\bf ZFC})   The cardinal  of the  largest thin $\Pi_1^1$-set in  $\Sio$ is equal to the cardinal of  $\om_1^{\bf L}$. 
\end{theorem}

This means that in a given model {\bf V} of {\bf ZFC} the cardinal  of the  largest thin $\Pi_1^1$-set in  $\Sio$ is equal to the cardinal 
{\it in {\bf V}} of  $\om_1^{\bf L}$, the ordinal which plays the role of the cardinal $\aleph_1$ in the inner model {\bf L}  of constructible sets of {\bf V}.

 We can now state the following theorem which will be useful in the sequel.  It follows from  Theorem \ref{card} and from some constructions 
of models of set theory due to Cohen (for (a)), Levy (for (b)) and Cohen (for (c)), see   \cite{Jech}. 

\begin{theorem}\label{cor1}
 \hs 
(a)   There is a model ${\bf V}_1$ of {\bf ZFC} in which the largest thin $\Pi_1^1$-set in  $\Sio$ has cardinal $\aleph_1$ with 
 $\aleph_1=2^{\aleph_0}$. 

(b)     There is a model ${\bf V}_2$ of {\bf ZFC} in which the largest thin $\Pi_1^1$-set in  $\Sio$ has cardinal $\aleph_0$, 
i.e.  is countable. 

(c) There is a model ${\bf V}_3$ of {\bf ZFC} in which the largest thin $\Pi_1^1$-set in  $\Sio$ has cardinal $\aleph_1$ with 
$\aleph_0 < \aleph_1 < 2^{\aleph_0}$. 

\end{theorem}

\noi  In particular, all models of  ({\bf ZFC + V=L}) satisfy  (a).  The  models of {\bf ZFC}
satisfying   (b)  are the models of  ({\bf ZFC} + $\om_1^{\bf L} < \om_1$).

\section{Cardinality problems for $\om$-languages}

\begin{theorem}\label{mainthe}
There exists a  real-time $1$-counter B\"uchi automaton $\mathcal{A}$ such that the  cardinality of the complement $L(\mathcal{A})^-$ of the 
$\om$-language $L(\mathcal{A})$ is not determined by the axiomatic system {\bf ZFC}: 

 (1). 
There is a model $V_1$ of  {\bf ZFC} in which      $L(\mathcal{A})^-$ is countable. 

(2). 
There is a model $V_2$ of  {\bf ZFC} in which     $L(\mathcal{A})^-$ has cardinal $2^{\aleph_0}$. 

(3). 
There is a model $V_3$ of  {\bf ZFC} in which      $L(\mathcal{A})^-$ has cardinal $\aleph_1$ with 
$\aleph_0<\aleph_1<2^{\aleph_0}$.

\end{theorem}

\proo   
 From now on we set   $\Si=\{0, 1\}$  and      we   shall  denote by $\mathcal{C}_1$ 
the largest thin $\Pi_1^1$-set in $\{0, 1\}^\om=2^\om$.  

This set  $\mathcal{C}_1$  is a $\Pi_1^1$-set  defined by a $\Pi_1^1$-formula $\phi$, given by 
Moschovakis  in   \cite[page 248]{Moschovakis80}.  
Thus its complement 
$\mathcal{C}_1^-=2^\om - \mathcal{C}_1$ is a   $\Si_1^1$-set  defined by the  $\Si_1^1$-formula $\psi=\neg \phi$. 
By Proposition \ref{tm},  the $\om$-language  $\mathcal{C}_1^-$ is accepted by a B\"uchi Turing machine $\mathcal{M}$ and by a  
 $2$-counter  B\"uchi  automaton $\mathcal{A}_1$ which can be effectively constructed. 

 We are now going to use some constructions  which were  used in a previous paper \cite{Fin-mscs06} to study topological properties of context-free 
$\om$-languages, and which will be useful in the sequel. 

 Let   $E$ be a new letter not in 
$\Si$,  $S$ be an integer $\geq 1$, and $\theta_S: \Sio \ra (\Sigma \cup \{E\})^\om$ be the 
function defined, for all  $x \in \Sio$, by: 
$$ \theta_S(x)=x(1).E^{S}.x(2).E^{S^2}.x(3).E^{S^3}.x(4) \ldots 
x(n).E^{S^n}.x(n+1).E^{S^{n+1}} \ldots $$

We proved in \cite{Fin-mscs06} that if    $L \subseteq \Sio$ is an 
$\om$-language in the class  ${\bf BCL}(2)_\om$ and   $k=cardinal(\Si)+2$, $S=(3k)^3$, then one can effectively construct  from 
a    B\"uchi $2$-counter automaton  $\mathcal{A}_1$  accepting $L$                a real time 
B\"uchi $8$-counter automaton $\mathcal{A}_2$ such that $L(\mathcal{A}_2)=\theta_S(L)$. 

 On the other hand, it is easy to see that $\theta_S(\Sio)^-=(\Sigma \cup \{E\})^\om - \theta_S(\Sio)$ is accepted 
by a real time B\"uchi $1$-counter automaton. The class 
{\bf r}-${\bf BCL}(8)_\om \supseteq$ {\bf r}-${\bf BCL}(1)_\om$  is closed under  finite union in an effective way, so 
$\theta_S(L) \cup    \theta_S(\Sio)^-$ is accepted by a real time B\"uchi $8$-counter automaton $\mathcal{A}_3$ which can be effectively constructed 
from     $\mathcal{A}_2$. 

   In \cite{Fin-mscs06} we used also another coding which we now recall. 
Let    $K = 2 \times 3 \times 5 \times 7 \times 11 \times 13 \times 17 \times 19 = 9699690$       be       the product of the eight first prime numbers. 
Let $\Ga$ be a finite alphabet; here we shall set $\Ga=\Si\cup \{E\}$.  An $\om$-word $x\in \Gao$ is coded by the $\om$-word 
$$h_K(x)=A.C^K.x(1).B.C^{K^2}.A.C^{K^2}.x(2).B.C^{K^3}.A.C^{K^3}.x(3).B \ldots  
B.C^{K^n}.A.C^{K^n}.x(n).B \ldots  $$

\noi over the alphabet $\Ga_1=\Ga \cup \{A, B, C\}$, where $A, B, C$ are new letters not in $\Ga$. 
We  proved in \cite{Fin-mscs06} that, from a  real time B\"uchi $8$-counter automaton $\mathcal{A}_3$ accepting $L(\mathcal{A}_3) \subseteq \Gao$, 
one can effectively construct a  B\"uchi $1$-counter automaton $\mathcal{A}_4$ accepting the $\om$-language 
$h_K( L(\mathcal{A}_3) )$$ \cup h_K(\Ga^{\om})^-$. 

 Consider  now  the mapping 
 $\phi_K: (\Ga \cup\{A, B, C\})^\om \ra (\Ga \cup\{A, B, C,  F\})^\om $ which is simply defined by:  
for all $x\in (\Ga \cup\{A, B, C\})^\om$, 
$$\phi_K(x) = F^{K-1}.x(1).F^{K-1}.x(2)  
\ldots F^{K-1}.x(n). F^{K-1}.x(n+1).F^{K-1} \ldots$$

\noi Then the $\om$-language 
$\phi_K(L(\mathcal{A}_4))=\phi_K ( h_K( L(\mathcal{A}_3) )$$ \cup h_K(\Ga^{\om})^- )$ is accepted by  
 a  real time B\"uchi $1$-counter automaton $\mathcal{A}_5$ which can be effectively 
constructed from the B\"uchi $8$-counter automaton $\mathcal{A}_4$, \cite{Fin-mscs06}. 

 On the other hand,    
it is easy to see that the $\om$-language $ (\Ga \cup\{A, B, C, F\})^\om  - \phi_K( (\Ga \cup\{A, B, C\})^\om )$ is $\om$-regular 
and to construct a ($1$-counter) B\"uchi automaton 
accepting it. Then one can effectively construct from $\mathcal{A}_5$ a real time B\"uchi $1$-counter automaton $\mathcal{A}_6$ 
accepting the $\om$-language 
$\phi_K ( h_K( L(\mathcal{A}_3) )$$ \cup h_K(\Ga^{\om})^- ) \cup \phi_K( (\Ga \cup\{A, B, C\})^\om )^-$.

To sum up:  we have obtained, from a B\"uchi Turing machine $\mathcal{M}$ accepting the $\om$-language 
$\mathcal{C}_1^-\subseteq \Sio=2^\om$, a $2$-counter  B\"uchi  automaton $\mathcal{A}_1$ accepting the same $\om$-language, 
a real time B\"uchi $8$-counter automaton $\mathcal{A}_3$ accepting the $\om$-language  
$L(\mathcal{A}_3)=\theta_S(\mathcal{C}_1^-) \cup    \theta_S(\Sio)^-$, 
 a B\"uchi $1$-counter automaton $\mathcal{A}_4$ accepting the $\om$-language  $h_K( L(\mathcal{A}_3) )$$ \cup h_K(\Ga^{\om})^-$, and 
a real time B\"uchi $1$-counter automaton $\mathcal{A}_6$ accepting the $\om$-language 
$\phi_K ( h_K( L(\mathcal{A}_3) )$$ \cup h_K(\Ga^{\om})^- ) \cup \phi_K( (\Ga \cup\{A, B, C\})^\om )^-$. 
From now on we shall  denote simply $\mathcal{A}_6$ by $\mathcal{A}$.  

 Therefore  we have successively the following equalities: 

 $L(\mathcal{A}_1)=\mathcal{C}_1^-$, 

 $L(\mathcal{A}_1)^-=\mathcal{C}_1$, 

$L(\mathcal{A}_3)^-=\theta_S(\mathcal{C}_1)$, 

 $L(\mathcal{A}_4)^-=h_K( L(\mathcal{A}_3) ^-)=h_K(\theta_S(\mathcal{C}_1))$, 

 $L(\mathcal{A}_6)^-=\phi_K(h_K( L(\mathcal{A}_3) ^-))=\phi_K(h_K( \theta_S(\mathcal{C}_1)))$. 

 This implies easily that the $\om$-languages $L(\mathcal{A}_1)^-$, $L(\mathcal{A}_3)^-$, $L(\mathcal{A}_4)^-$, and 
$L(\mathcal{A}_6)^-=L(\mathcal{A})^-$ all 
have the same cardinality as the set $\mathcal{C}_1$, because each of the maps $\theta_S$, $h_K$ and $\phi_K$ is injective. 

 Thus we can infer the result stated in the theorem from the above Theorem \ref{cor1}. 
\ep 

\hs The following corollary follows directly from Item (3) of Theorem \ref{mainthe}. 

\begin{Cor}
It is consistent with {\bf ZFC} that the Continuum Hypothesis is not satisfied for complements  of  $1$-counter $\om$-languages, (hence also 
for  complements of context-free $\om$-languages). 
\end{Cor}

\begin{Rem}
This can be compared with the fact that the Continuum Hypothesis is satisfied for regular languages of infinite trees  (which are closed under complementation), 
 proved by  Niwinski  in \cite{Niwinski91}. Notice that this may seem  amazing because 
from a topological point of view one can find  regular tree languages
which are more complex than  context-free $\om$-languages, as there are  
regular tree languages in the class ${\Delta}^1_2 \setminus {\bf \Sigma}^1_1 \cap {\bf \Pi}^1_1$ 
while context-free $\om$-languages are all analytic, i.e. ${\bf \Sigma}^1_1$-sets.  
\end{Rem}

  Recall that a  real-time $1$-counter B\"uchi automaton  $\mathcal{C}$ has a finite description to which can be associated, in an effective way,  
 a unique natural number called the index of 
$\mathcal{C}$. 
From now on, we shall  denote, as in \cite{Fin-HI}, by $\mathcal{C}_z$ the  real time B\"uchi $1$-counter automaton of index $z$ 
(reading words over $\Omega=\{0, 1,  A, B, C, E, F\}$).

 We can now use the  proofs of Theorem \ref{cor1} and \ref{mainthe} to prove  that some natural cardinality problems  
 are actually located at the {\bf third level} of the analytical hierarchy. 
The notions  of analytical hierarchy on subsets of $\mathbb{N}$ 
and of  classes of this hierarchy may be found for instance in \cite{cc} or in  the textbook  \cite{rog}.

\begin{theorem}
\hs
(1). 
$\{  z \in \mathbb{N}  \mid  L(\mathcal{C}_z)^- \mbox{ is finite } \}$ is $\Pi_2^1$-complete. 

(2). 
$\{  z \in \mathbb{N}  \mid  L(\mathcal{C}_z)^- \mbox{ is countable } \}$ is in $\Si_3^1 \setminus (\Pi_2^1 \cup \Si_2^1)$. 

(3).
 $\{  z \in \mathbb{N}  \mid  L(\mathcal{C}_z)^- \mbox{ is uncountable}  \}$ is in $\Pi_3^1 \setminus (\Pi_2^1 \cup \Si_2^1)$. 

\end{theorem}

\proo  Item (1) was proved in \cite{Fin-HI}, and  item (3)  follows directly from item (2).  

 We now   prove  item (2).  We first show that $\{  z \in \mathbb{N}  \mid  L(\mathcal{C}_z)^- \mbox{ is countable } \}$ is in the class $\Si_3^1$. 

 Notice first that, using a recursive bijection $b: (\mathbb{N}^\star)^2 \ra \mathbb{N}^\star$, 
 we can consider an infinite word over a finite alphabet $\Omega$ as a countably infinite family of infinite words over the same 
alphabet by considering, for any $\om$-word $\sigma \in \Omega^\om$, the family of $\om$-words 
$(\sigma_i)_{i \geq 1}$ such that for each $i \geq 1$ 
 the $\om$-word $\sigma_i \in \Omega^\om$ is defined by $\sigma_i(j)= \sigma(b(i, j))$ for each $j\geq 1$. 

 We can  now express ``$L(\mathcal{C}_z)^- \mbox{ is countable }$" by the formula: 
$$\exists \sigma \in \Omega^\om ~~ \fa x \in \Omega^\om ~[ ( x \in L(\mathcal{C}_z)  ) \mbox{ or } (  \exists i \in \mathbb{N} ~ x = \sigma_i ) ]$$

 This is a $\Si_3^1$-formula because $``( x \in L(\mathcal{C}_z) )"$, and hence also 
$``[ ( x \in L(\mathcal{C}_z)  ) \mbox{ or } (  \exists i \in \mathbb{N} ~ x= \sigma_i  ) ]"$,  are  expressed by  $\Si_1^1$-formulas. 

 We can now prove that $\{  z \in \mathbb{N}  \mid  L(\mathcal{C}_z)^- \mbox{ is countable } \}$ is neither in the class $ \Si_2^1$ nor in the 
class $\Pi_2^1$, by using Shoenfield's  Absoluteness  Theorem from Set Theory. 

 Let   $\mathcal{A}$ be the real-time $1$-counter B\"uchi automaton cited in Theorem \ref{mainthe} and let $z_0$ be its index 
so that $\mathcal{A}=\mathcal{C}_{z_0}$.  
Assume   that {\bf V} is  a model of ({\bf ZFC} + $\om_1^{\bf L} < \om_1$).   In the model {\bf V},   the integer $z_0$ belongs to the set 
$ \{  z \in \mathbb{N}  \mid L(\mathcal{C}_z)^- \mbox{ is countable } \}$, while in 
 the inner model     ${\bf L } \subseteq {\bf V }$, the language $L(\mathcal{C}_{z_0})^-$ has 
the cardinality of the continuum: 
thus  in  {\bf L }     the integer $z_0$ does not belong to the set 
$ \{  z \in \mathbb{N}  \mid   L(\mathcal{C}_z)^- \mbox{ is countable }  \}$. 
On the other hand, Shoenfield's Absoluteness Theorem implies that every  $\Si_2^1$-set (respectively,  $\Pi_2^1$-set)
 is absolute for all inner models of {\rm  (ZFC)}, 
see \cite[page 490]{Jech}.
 In particular, if the set   $ \{  z \in \mathbb{N}  \mid   L(\mathcal{C}_z)^- \mbox{ is countable } \}$   was a  $\Si_2^1$-set or a $\Pi_2^1$-set
 then it could not be a different subset of $\mathbb{N}$ in the models  ${\bf  V}$  and   ${\bf L }$ considered above.
Therefore, 
the   set  $ \{  z \in \mathbb{N}  \mid  L(\mathcal{C}_z)^- \mbox{ is countable }  \}$   is neither  a $\Si_2^1$-set nor a $\Pi_2^1$-set. 
\ep

\begin{Rem}
Using an easy coding we can obtain a similar result for $1$-counter automata reading words over  $\Si$, where $\Si$
is any finite alphabet having at least two letters. 
\end{Rem}

Notice that the same proof gives a partial answer to a question of Castro and Cucker. They stated in \cite{cc} that 
the problem to determine whether the complement of the $\om$-language accepted by a given Turing machine is countable  (respectively, uncountable) is 
in the class $\Si_3^1$ (respectively, $\Pi_3^1$), and  asked 
for the exact complexity of these decision problems. 

\begin{theorem}
The problem 
to determine whether the complement of the $\om$-language accepted by a given Turing machine is countable  (respectively, uncountable) is in the class 
 $\Si_3^1 \setminus (\Pi_2^1 \cup \Si_2^1)$        (respectively,     $\Pi_3^1 \setminus (\Pi_2^1 \cup \Si_2^1)$). 
\end{theorem}

 We now consider  acceptance of binary  relations over infinite words by $2$-tape B\"uchi automata, firstly considered by 
Gire and Nivat in \cite{Gire-Nivat}.  A $2$-tape  automaton is an automaton having two tapes and two reading heads, one for each tape, 
which can move asynchronously, and a finite control as in the case of a ($1$-tape) automaton. The automaton reads a pair of (infinite) words 
$(u, v)$ where $u$ is on the first tape and $v$ is on the second tape, so that  a $2$-tape B\"uchi automaton $\mathcal{B}$ 
accepts an  infinitary rational relation $L(\mathcal{B})\subseteq \Si_1^\om \times \Si_2^\om$, where $\Si_1$ and $\Si_2$ are two finite alphabets. 
Notice that  $L(\mathcal{B})\subseteq \Si_1^\om \times \Si_2^\om$ may be seen as  an $\om$-language over the product alphabet 
$\Si_1 \times \Si_2$.

\hs We shall use  a coding used in a 
previous paper \cite{Fin06b} on the topological complexity of infinitary rational relations. 
We first  recall  a coding of  an $\om$-word over the finite alphabet $\Omega=\{0, 1,  A, B, C, E, F\}$
by an  $\om$-word over the  alphabet $\Omega' = \Omega \cup \{D\}$, where  $D$ is an additionnal letter 
not in $\Omega$. 
 For $x\in \Omega^\om$  the $\om$-word $h(x)$ is defined by : 
$$h(x) = D.0.x(1).D.0^2.x(2).D.0^3.x(3).D \ldots D.0^n.x(n).D.0^{n+1}.x(n+1).D \ldots$$
\noi It is easy to see that the mapping $h$ from $\Omega^\om$ into $(\Omega \cup \{D\})^\om$ is injective. 
 Let now  $\alpha$ be the $\om$-word over the alphabet $\Omega'$ 
 which is simply defined by:
$$\alpha = D.0.D.0^2.D.0^3.D.0^4.D \ldots D.0^n.D.0^{n+1}.D \ldots$$
\noi The following result  was  proved in \cite{Fin06b}.

\begin{Pro}[\cite{Fin06b}]\label{pro-ratrel} 
 Let  $L \subseteq \Omega^\om$ be in  {\bf r}-${\bf BCL}(1)_\om$ and $\mathcal{L}= h(L)  \cup (h(\Omega^{\om}))^- $.  
 Then 
 $$R = \mathcal{L} \times \{\alpha\} ~~ \bigcup  ~~(\Omega')^\om \times ( (\Omega')^\om - \{\alpha\})$$
\noi is an  infinitary rational relation. 
 Moreover  one can effectively construct  from a real time $1$-counter B\"uchi automaton $\mathcal{A}$ accepting $L$  
a $2$-tape B\"uchi automaton $\mathcal{B}$ accepting the infinitary relation $R$. 
\end{Pro}

 We can now prove our second main  result. 

\begin{theorem}\label{mainthe2}
There exists a  $2$-tape  B\"uchi automaton $\mathcal{B}$ such that the  cardinality of the complement of the infinitary rational relation
 $L(\mathcal{B})$ is not determined by  {\bf ZFC}. Indeed it holds that: 

(1).  There is a model $V_1$ of  {\bf ZFC} in which      $L(\mathcal{B})^-$ is countable. 

(2).  There is a model $V_2$ of  {\bf ZFC} in which     $L(\mathcal{B})^-$ has cardinal $2^{\aleph_0}$. 

(3).  There is a model $V_3$ of  {\bf ZFC} in which      $L(\mathcal{B})^-$ has cardinal $\aleph_1$ with 
$\aleph_0<\aleph_1<2^{\aleph_0}$.

\end{theorem}

\proo  Let   $\mathcal{A}$ be the real time $1$-counter B\"uchi automaton constructed  in the proof of  Theorem \ref{mainthe},  and 
$\mathcal{B}$ be the $2$-tape B\"uchi automaton which can be constructed from  $\mathcal{A}$ by the above Proposition \ref{pro-ratrel}.
 Letting  $L=L(\mathcal{A})$,   the complement of the infinitary rational relation $R=L(\mathcal{B})$ is equal to  
~~~
$[  (\Omega \cup \{D\})^\om -   \mathcal{L} ]   \times \{\alpha\}= h( L^-) \times \{\alpha\}.$
~~ Thus the cardinality of $R^-=L(\mathcal{B})^-$ is equal to the  cardinality of the $\om$-language $h( L^-) $, so that the result follows  from
 Theorem \ref{mainthe}. 
\ep 

\hs As in the case of $\om$-languages of  $1$-counter automata, we can now state the following result, where $\mathcal{T}_z$ is the 
$2$-tape B\"uchi automaton of index $z$ reading words over $\Omega' \times \Omega'$. 

\begin{theorem}\label{undec}
\hs 

(1).
$\{  z \in \mathbb{N}  \mid  L(\mathcal{T}_z)^- \mbox{ is finite } \}$ is $\Pi_2^1$-complete. 

(2).
$\{  z \in \mathbb{N}  \mid  L(\mathcal{T}_z)^- \mbox{ is countable } \}$ is in $\Si_3^1 \setminus (\Pi_2^1 \cup \Si_2^1)$. 

(3).
 $\{  z \in \mathbb{N}  \mid  L(\mathcal{T}_z)^- \mbox{ is uncountable}  \}$ is in $\Pi_3^1 \setminus (\Pi_2^1 \cup \Si_2^1)$. 

\end{theorem}

\proo Item (1) was proved in \cite{Fin-HI}. Items (2) and (3) are proved  similarly to the case of $\om$-languages of  $1$-counter automata, using 
Shoenfield's Absoluteness Theorem. 
\ep 

\hs On the other hand we have the following result. 

\begin{Pro}\label{dec}
It is decidable whether an infinitary rational relation  $R \subseteq \Si_1^\om \times \Si_2^\om$, accepted by a given 
$2$-tape B\"uchi automaton $\mathcal{B}$, is countable (respectively, uncountable). 
\end{Pro}

\proo Let $R \subseteq \Si_1^\om \times \Si_2^\om$ be an infinitary rational relation accepted by a $2$-tape B\"uchi automaton $\mathcal{B}$. 
It is known that ${\rm Dom}(R)=\{ u\in \Si_1^\om \mid \exists v \in \Si_2^\om ~~ (u,v) \in R \}$ and 
${\rm Im}(R)=\{ v\in \Si_2^\om \mid \exists u \in \Si_1^\om ~~ (u,v) \in R \}$ are 
regular $\om$-languages and that one can find B\"uchi automata $\mathcal{A}$ and $\mathcal{A}'$ accepting ${\rm Dom}(R)$ and 
${\rm Im}(R)$, \cite{Gire-Nivat}.  On the other hand Lindner and Staiger have proved 
 that one can compute the cardinal of  a given regular $\om$-language $L(\mathcal{A})$  (see  \cite{KuskeLohrey} where Kuske and Lohrey 
proved that this problem is actually in the class PSPACE). But it is easy to see that the infinitary rational relation $R$ is countable if and only if the two 
$\om$-languages ${\rm Dom}(R)$ and ${\rm Im}(R)$ are countable, thus one can decide whether the infinitary rational relation $R$ is countable 
 (respectively, uncountable). 
\ep

\begin{Rem}
The results given by Items (2) and (3) of Theorem \ref{undec} and Proposition \ref{dec} are 
rather surprising: they show that there is a remarkable gap between the complexity of the same decision problems 
for infinitary rational relations and for their complements, as  there 
is a big space between the class  $\Delta_1^0$ of computable  sets and the class  $\Si_3^1 \setminus (\Pi_2^1 \cup \Si_2^1)$.
\end{Rem}

\section{Concluding remarks}

~~~~~We have proved that amazingly some  basic cardinality 
questions on automata reading infinite words depend on the models of the axiomatic system 
{\bf ZFC}. 

 In \cite{Fin-ICST} we have proved that 
 the topological complexity of an $\om$-language accepted by a $1$-counter B\"uchi automaton, or of an infinitary rational relation 
accepted by a $2$-tape B\"uchi automaton, 
is not determined by  {\bf ZFC}. 

In \cite{jaf28}, we study some 
cardinality questions for B\"uchi-recognizable languages of infinite pictures and prove results which are   similar to those we 
have obtained in this paper for $1$-counter $\om$-languages and for infinitary rational relations. 

The next step in this research project would  be to determine which  properties of automata actually  depend on the  models of  
{\bf ZFC}, and to achieve a more complete investigation of these properties.

\section*{Annexe} 

\hs \noi {\bf Proof of Thorem \ref{cor1}.}

\hs   (a).  In the model {\bf L},  the cardinal of the  largest thin $\Pi_1^1$-set in  $\Sio$ is equal to the cardinal of  $\om_1$. 
Moreover the continuum hypothesis is satisfied thus $2^{\aleph_0}=\aleph_1$:  
 thus the largest thin $\Pi_1^1$-set in  $\Sio$ has the cardinality $2^{\aleph_0}=\aleph_1$.

\hs  (b).  Let {\bf V} be a model of  ({\bf ZFC} + $\om_1^{\bf L} < \om_1$).  Since   $\om_1$ is the first uncountable ordinal in  {\bf V}, 
$\om_1^{\bf L} < \om_1$ implies that $\om_1^{\bf L}$ is a countable ordinal in {\bf V}. Its cardinal is $\aleph_0$, and therefore this is also 
 the cardinal in    {\bf V}    of 
the  largest thin $\Pi_1^1$-set in  $\Sio$.  

\hs (c).  It suffices to show that there is a model ${\bf V}_3$ of {\bf ZFC} in which $\om_1^{\bf L} = \om_1$ and $\aleph_1 < 2^{\aleph_0}$. 
Such a model can be constructed by Cohen's forcing:  start  from a model 
{\bf V} of {\bf ZFC + V=L} (in which   $\om_1^{\bf L} = \om_1$)        and construct by  forcing  a generic extension {\bf V[G]} in which are added 
$\aleph_2$ (or even more) ``Cohen's reals",  which are in fact $\aleph_2$  subsets of $\om$. 
 Notice that the cardinals are preserved under this extension (see 
\cite[page 219]{Jech}),  and that the constructible sets of  {\bf V[G]} are also  the   constructible sets of    {\bf V},  
 thus in the new model {\bf V[G]} of    {\bf ZFC}     we still have 
$\om_1^{\bf L} = \om_1$,  but now $\aleph_1 < 2^{\aleph_0}$. 
\ep

\end{document}